\documentclass[doublecol]{epl2} 
\usepackage{amsmath}
\usepackage{graphicx}
\usepackage{bm}
\usepackage{amsfonts}

\title{Breakdown of a renormalized perturbation expansion around
mode-coupling theory of the glass transition}
\shorttitle{Breakdown of a renormalized perturbation expansion around MCT}

\author{Grzegorz Szamel\inst{1} \and Elijah Flenner\inst{1} \and Hisao Hayakawa\inst{2}}
\shortauthor{Grzegorz Szamel \etal}

\institute{
\inst{1} Department of Chemistry, 
Colorado State University, Fort Collins, CO 80523, USA \\
\inst{2} Yukawa Institute for Theoretical Physics, Kyoto University,
Kyoto 606-8502, Japan
}

\date{\today}


\abstract{
We analyze a renormalized perturbation expansion around the mode-coupling
theory of the glass transition. We focus on the long-time limit of the 
irreducible memory function. We discuss a renormalized diagrammatic
expansion for this function and re-sum two infinite classes of diagrams. 
We show that the resulting contributions to the irreducible memory function 
diverge at the mode-coupling transition. A further re-summation
of ladder diagrams constructed by iterating these divergent contributions
gives a finite result which cancels the mode-coupling theory's 
expression for the irreducible memory function.
}

\begin{document}

\maketitle

\section{Introduction}

Since its introduction almost thirty years ago, the mode-coupling theory (MCT) 
\cite{Leutheusser,BGS,DMRT,Goetzebook} has significantly contributed
to our understanding of the slowing down of a fluid's dynamics 
upon approaching the glass transition.
In particular, the theory accounts for the cage 
effect: in a super-cooled fluid a given particle spends considerable time in its 
solvation shell before making any further motion. This physical picture results in 
intermediate time plateaus in the mean-square displacement and 
the intermediate scattering function. Both the values of the plateaus 
of these functions and their time dependence in the region of the plateaus
are well described by 
the mode-coupling theory \cite{NK,KNS,GK2000,WPFV2010}. 
More generally, the theory accurately describes the initial stages of
the slowing down upon super-cooling. 
However, it also predicts a 
spurious dynamic transition, the so-called mode-coupling transition. 
It predicts that upon further super-cooling the time scale of the relaxation 
from the plateau region diverges
as a power law. 
Extensive simulational
studies showed that instead of this transition there is a smooth 
cross-over, with relaxation times and transport coefficients following 
mode-coupling-like power laws only for a few decades \cite{KobLH}. 

The microscopic derivation of the mode-coupling theory allows, in principle, for its
extension, which could address the above mentioned fundamental problem. 
Two such extensions were proposed shortly after the original theory was derived
\cite{DM,GScutoff}. More recently, these extensions were critically assessed 
and found somewhat inadequate \cite{CR}. In addition to the arguments presented
in Ref. \cite{CR}, 
the reason why neither of these extensions is satisfactory  
is that they rely upon couplings to current modes that are defined for systems
with Newtonian dynamics. Thus, these extensions cannot be applied to systems with
Brownian dynamics. In contrast, computer simulations showed that deviations
from mode-coupling-like power laws are qualitatively the same in systems with
both dynamics \cite{SE}.

In fact, most of our current understanding of the significance of
the processes neglected within the mode-coupling theory is indirect. 
Within the so-called Franz-Parisi potential approach (originally introduced 
in the context of mean-field spin glasses \cite{FP} but later generalized
to super-cooled fluids \cite{CFP}), the mode-coupling transition is identified
with the appearance of a metastable state with a non-zero value of the overlap
between the fluid and a template. 
This metastable state becomes the absolute
minimum of the 
Franz-Parisi potential only at the thermodynamic glass transition.
This approach suggests that the mode-coupling theory treats 
the metastable state as absolutely stable, thus 
neglecting two types of dynamic events. First, near the transition there should
be critical fluctuations. 
At the very least, these fluctuations should move the mode-coupling 
transition towards lower temperatures and/or higher densities. 
Second, if the transition survives the inclusion of the 
fluctuations, it should be cut off by activated events, which 
are sometimes referred to as ``hopping processes''. 
This general physical picture is supported by results obtained
for spin glass models with long-but-finite range interactions \cite{Franz,FM}.
However, explicit calculations on 
particle-based models have been 
lacking\footnote{The activated barrier hopping theory of 
Schweizer and Saltzman \cite{Ken} is a step in 
this direction. It is, however, restricted to single particle dynamics. In addition,  
its relation with the Franz-Parisi potential picture is unclear.}
until the recent analysis of critical fluctuations 
\cite{Fetal}. 
This analysis is based on a static replica field theory approach. 

The main problem with extending the dynamic approach beyond the 
mode-coupling theory comes from the original, projection operator 
derivation of the theory \cite{Goetzebook}. This derivation is rather opaque and, 
therefore, it is difficult to 
generalize\footnote{The exception is the so-called generalized mode-coupling theory
proposed by one of us \cite{gMCT} and subsequently further developed by 
Wu and Cao \cite{WC} and Mayer \textit{et al.} \cite{Mayer}.}.
A more promising avenue is to start from a diagrammatic expansion  
and then resort to re-summations. 
However, most diagrammatic expansions 
derived to date \cite{ABL,KK,NH,JW} are quite complicated and, therefore, are
unlikely to produce results going beyond the mode-coupling theory.

Recently, one of us developed an alternative diagrammatic approach \cite{GSdiagram}.
To make some technical steps easier, it was assumed  
that the microscopic dynamics is Brownian. The same assumption will be used in
this Letter. 
The starting point of the derivation presented in
Ref. \cite{GSdiagram} is the hierarchy of equations of motion for 
correlation functions of many-particle densities orthogonalized 
with respect to the equilibrium probability distribution. Using orthogonalized 
densities results in two advantages. First, bare inter-particle interactions get
replaced by renormalized interactions, which can be expressed in terms of the 
derivatives of the equilibrium correlation functions. Second, the initial condition for 
the set of correlation functions of orthogonalized densities is very simple,
which simplifies the structure of the diagrams. The second step
of the derivation is a perturbative solution of the hierarchy of equations of motion. 
The terms in the perturbation expansion are represented by diagrams. After the
expansion is derived, one can use all standard diagrammatic techniques including 
re-summations and Dyson equation-type analyses. 
Within this approach, 
the mode-coupling theory amounts to a self-consistent one-loop 
approximation for the so-called irreducible memory function \cite{CHess}. 
Notably, the structure of
the diagrammatic expansion of Ref. \cite{GSdiagram} is relatively simple. This 
made it possible to use this approach to show that a four-point dynamic density
correlation function contains a divergent contribution \cite{GS4point}
and to evaluate two corrections to the mode-coupling expression for 
the long-time limit of the 
memory function \cite{GSPTEP}.

Here, we significantly extend this last contribution. 
Again, we focus on the long-time limit
of the irreducible memory function. Thus, like in static approaches we deal with 
time-independent quantities. However, in contrast to the latter approaches,
our theory is derived from dynamics. We use the new, renormalized diagrammatic expansion
for the memory function, which was suggested in Ref. \cite{GSPTEP}. We discuss
two infinite classes of renormalized diagrams, which have 
a clear physical interpretation. After re-summation, these two classes of diagrams 
result in corrections to the mode-coupling approximation that diverge upon
approaching the mode-coupling transition in dimensions $D<4$.
We note that a subsequent re-summation of a whole series of divergent contributions
produces a result, which is finite at the mode-coupling transition, but
cancels the original mode-coupling contribution to the irreducible memory function.
Our findings suggest a breakdown of a renormalized perturbation expansion 
around the mode-coupling theory. 

\section{Diagrammatic expansion for the irreducible
memory function}
Here, we discuss two different expansions for the irreducible memory 
function. We start with diagrams introduced in Ref. \cite{GSdiagram}.
Briefly, in these diagrams, bonds represent bare propagation of density
fluctuations, vertices represent renormalized interactions and 
diagrams with an odd number of four-leg vertices contribute with a negative sign 
indicated next to the diagram. The bonds are defined only for
positive times. The time direction is indicated by an arrow attached to a bond.

\begin{figure}
\centerline{\includegraphics[scale=.135]{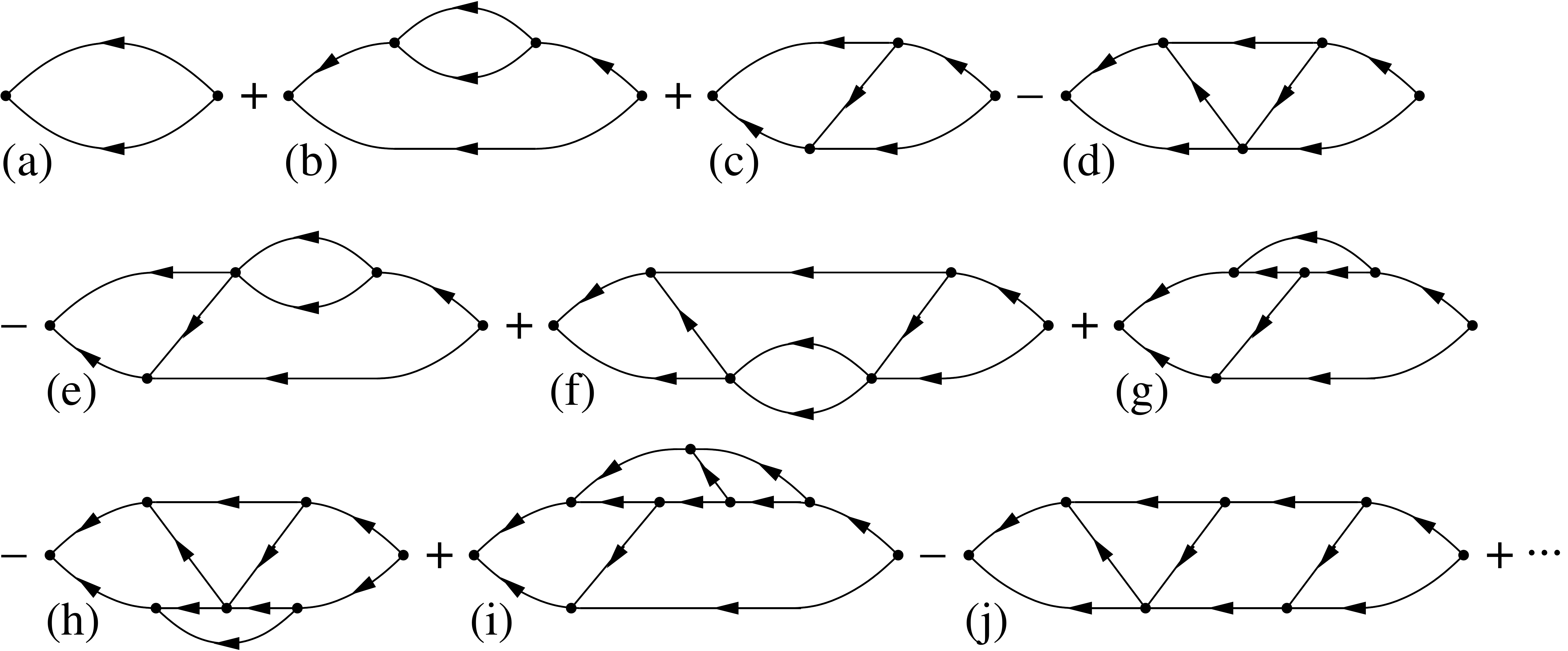}}
\caption{The first few diagrams contributing to the irreducible
memory function $M^{\text{irr}}(k;t)$.}
\label{fig1}
\end{figure}
The irreducible memory function $M^{\text{irr}}(k;t)$ contains
all non-trivial information about the dynamics of the system. Using 
a projection operator approach one can derive an exact but formal equation 
for $M^{\text{irr}}$, which involves the so-called irreducible evolution 
operator that has single particle dynamics projected out \cite{CHess}. 
It was showed in Ref. \cite{GSdiagram} that the 
irreducible memory function is represented by a sum of all diagrams 
that start with the right vertex and
end with the left vertex, and are one-particle irreducible, \textit{i.e.} they do
not separate into disconnected components upon removal of a single bond or a single 
four-leg vertex. The first few diagrams contributing to the irreducible
memory function are showed in Fig. \ref{fig1}. Furthermore, it was showed in 
Ref. \cite{GSdiagram} that the mode-coupling approximation amounts
to including only those diagrams 
that separate into two disconnected components upon removing the
left and right vertices. Thus, 
out of diagrams
showed in Fig. \ref{fig1} one keeps only diagrams (a-b). 
After all such diagrams are re-summed, one gets a diagram
whose topology is identical to that of diagram (a) in Fig. \ref{fig1}, but with
bare propagators replaced with the full propagators. 

From the point of view of the topology of the diagrams, the simplest 
diagrams neglected in the mode-coupling approximation\footnote{
These diagrams 
originate from couplings between different dynamic
modes that are neglected in the mode-coupling theory. For brevity we will 
refer to them as non-mode-coupling diagrams.}
are the diagrams that separate into disconnected
components upon removing the left and right vertices, and subsequent removing
of a single propagator or a single four-leg vertex, and which satisfy the 
following condition: each of the components should be one 
of the diagrams included in the mode-coupling approximation. 
Diagrams (c-e) and (g-h)
in Fig. \ref{fig1} belong to this class. 
Diagram (f) separates into disconnected
components upon removing the left and right vertices, and removing
of two successive four-leg vertices. We will argue below that this diagram
needs to be included together with the simplest non-mode-coupling
diagrams. In contrast, 
diagrams (i-j) in Fig. \ref{fig1} 
are not the simplest non-mode-coupling diagrams. 
Diagram (i) separates into 
two 
components upon removing the
left and right vertices, and removing of a single propagator,
but one of the resulting components is a diagram not included in the mode-coupling
approximation. 
Diagram (j) separates into disconnected
components upon removing the left and right vertices, and removing 
both a four-leg vertex and a single propagator. 

The mode-coupling expression for the memory function has the following important 
property: if the full propagator has a non-zero long-time limit, the memory function
also has a finite, non-zero long-time limit.
This is consistent with the exact 
equation expressing the time-derivative of intermediate 
scattering function $F(k;t)$ in terms of $M^{\text{irr}}(k;t)$ \cite{GSdiagram,SL,CHess}.
Namely, it follows from this equation that if the 
scattering
function has a non-zero limit, $\lim_{t\to\infty} F(k;t)=F(k)>0$, then
the irreducible 
memory function also has a non-zero limit,
$\lim_{t\to\infty} M^{\text{irr}}(k;t)=(D_0 k^2/S(k)) m(k)>0$,
where $S(k)$ is the static structure factor and $D_0$ is the
diffusion coefficient of an isolated Brownian particle, and
$F(k)$ and $m(k)$ are related by the following equation, 
$F(k)/(S(k)-F(k)) = m(k)$.
We shall emphasize that the only assumption used in writing this equation is that 
both the intermediate scattering function and the irreducible memory function have finite
non-zero limits. To simplify the nomenclature we will hereafter refer to $m(k)$ as
the long time limit of the irreducible memory function. 

\begin{figure}
\centerline{\includegraphics[scale=.135]{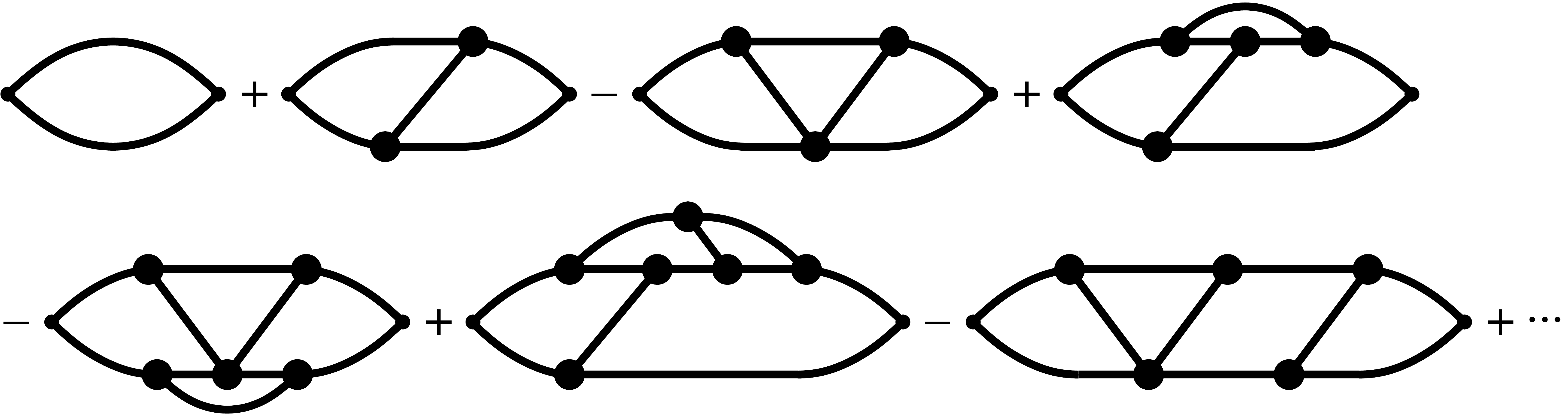}}
\caption{The first few renormalized 
diagrams contributing to the long-time limit of the irreducible
memory function $m(k)$.}
\label{fig2}
\end{figure}
As discussed in Ref. \cite{GSPTEP}, when calculating non-mode-coupling contributions
to $m(k)$ one has to make sure that the above discussed consistency is 
maintained. Thus, \textit{e.g.}, one cannot naively replace bare propagators by
full propagators in diagram (d) in Fig. \ref{fig1}, but one has to also include
diagram (f) and a whole class of similar diagrams. In this way one 
gets a new diagrammatic expansion in which all diagrams remain finite even if $F(k;t)$
has a non-zero long-time limit. The first few diagrams of the resulting expansion 
for $m(k)$ are shown in Fig. \ref{fig2}. In these diagrams bonds 
\includegraphics[scale=.14]{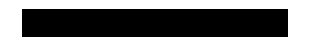} represent 
the long-time limit of the full intermediate scattering function,
$F(k)$, the right outside vertex 
\includegraphics[scale=.14]{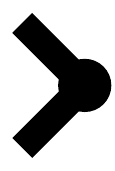} represents 
$n\tilde{v}_{\mathbf{k}_1+\mathbf{k}_2}
(\mathbf{k}_1,\mathbf{k}_2)$ where $n$ is the number
density and $\tilde{v}_{\mathbf{k}_1+\mathbf{k}_2}(\mathbf{k}_1,\mathbf{k}_2)=
(\mathbf{k}_1+\mathbf{k}_2)\cdot \left(c(k_1)\mathbf{k}_1+c(k_2)\mathbf{k}_2\right)
/|\mathbf{k}_1+\mathbf{k}_2|^2$ where $c(k)$ is the direct correlation function, 
the left outside vertex 
\includegraphics[scale=.14]{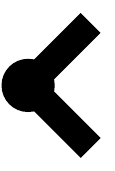} represents 
$S(|\mathbf{k}_1+\mathbf{k}_2|)\tilde{v}_{\mathbf{k}_1+\mathbf{k}_2}
(\mathbf{k}_1,\mathbf{k}_2)$, the right vertex inside the diagram 
\includegraphics[scale=.14]{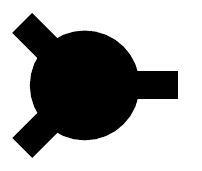} represents
$n\tilde{v}_{\mathbf{k}_1+\mathbf{k}_2}(\mathbf{k}_1,\mathbf{k}_2)
/m(|\mathbf{k}_1+\mathbf{k}_2|)$ and the left vertex inside the diagram 
\includegraphics[scale=.14]{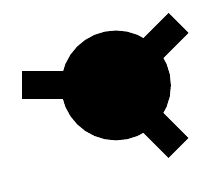} represents
$\tilde{v}_{\mathbf{k}_1+\mathbf{k}_2}(\mathbf{k}_1,\mathbf{k}_2)
/m(|\mathbf{k}_1+\mathbf{k}_2|)$. Finally, the four-leg vertex 
\includegraphics[scale=.14]{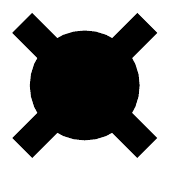} is a sum of two parts.
The more compact part reads $nS(|\mathbf{k}_1+\mathbf{k}_2|)
\tilde{v}_{\mathbf{k}_1+\mathbf{k}_2}(\mathbf{k}_1,\mathbf{k}_2)
\tilde{v}_{\mathbf{k}_3+\mathbf{k}_4}(\mathbf{k}_3,\mathbf{k}_4)
/m(|\mathbf{k}_1+\mathbf{k}_2|)$. The second part will be presented elsewhere 
\cite{SFH}. 

Diagrams contributing to $m(k)$ in the expansion showed in Fig. \ref{fig2}   
have the following property: one cannot cut a part out of these diagrams
by removing two bonds, one bond and one four-leg vertex, or two four-leg vertices.

\section{Divergent corrections to the 
mode-coupling contribution} 

The first diagram in Fig. \ref{fig2} represents the mode-coupling contribution
to $m(k)$. The contributions represented by the second
and third diagrams were calculated in Ref. \cite{GSPTEP}. In that calculation,
in the spirit of a perturbative expansion around mode-coupling theory, 
the exact long-time limits of the full scattering and memory functions, $F(k)$
and $m(k)$, were replaced by mode coupling approximations for these functions.  
It was showed that these two diagrams make a non-negligible contribution, 
which is negative at the mode-coupling transition.

\begin{figure}
\centerline{\includegraphics[scale=.135]{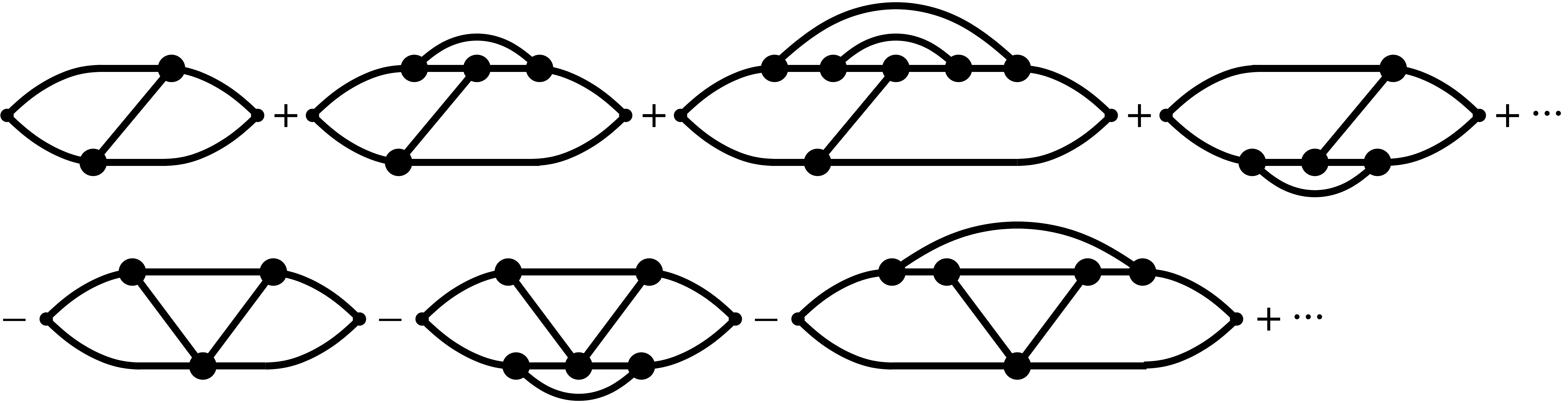}}
\caption{First few diagrams from two infinite classes of renormalized
diagrams contributing to the long-time limit of the irreducible memory function $m(k)$. 
Re-summation of these classes of diagrams results in two
diverging corrections to the mode-coupling approximation for $m(k)$.}
\label{fig3}
\end{figure}
Here we re-sum two infinite classes of diagrams contributing to $m(k)$.
Again, in the spirit of a perturbative expansion around mode-coupling theory, 
the exact long-time limits of the full scattering and memory functions
will be replaced by mode coupling approximations for these functions.
The two classes of diagrams are showed in Fig. \ref{fig3}.
The original motivation for considering these diagrams comes from the fact
that they originate from the simplest non-mode-coupling 
diagrams discussed in the preceding section (diagrams (d-h) in Fig. \ref{fig1}). 
Each of these diagrams separates into disconnected components
upon removing the left and right vertices and subsequently removing
one bond or one four-leg vertex. In turn, each of the resulting 
components has a characteristic rainbow-like insertion. 

\begin{figure}
\centerline{\includegraphics[scale=.135]{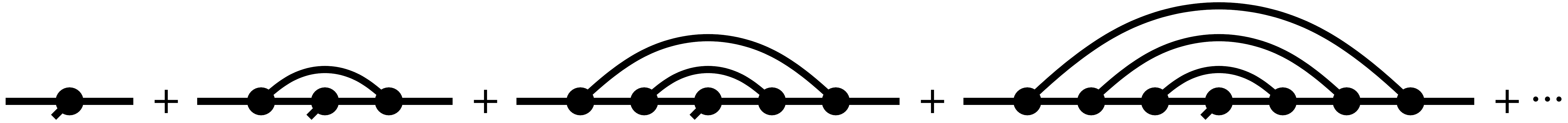}}
\caption{The first few rainbow diagrams. 
Each of the first 
three diagrams showed in this figure can be cut out of one of the first three
diagrams showed in Fig. \ref{fig3}.  
Re-summing the rainbow diagrams is
equivalent to solving linear integral equation (\ref{chiqk}).}
\label{fig4}
\end{figure}
To show that two classes of diagrams presented in Fig. \ref{fig3} can be re-summed 
we start from the analysis of rainbow diagrams showed in Fig. \ref{fig4}.
If we take $\mathbf{k}+\mathbf{q}/2$ as the incoming wave-vector,
$\mathbf{k}-\mathbf{q}/2$ as the outgoing one and 
$\mathbf{q}$ as a side wave-vector, we see that 
the sum of the rainbow diagrams showed in Fig. \ref{fig4}, which will be denoted by
$\chi_{\mathbf{q}}(\mathbf{k})$, satisfies the following linear integral equation,
\begin{eqnarray}\label{chiqk}
\chi_{\mathbf{q}}(\mathbf{k}) = \chi_{\mathbf{q}}^0(\mathbf{k})
+ \int \frac{d\mathbf{k}_1}{(2\pi)^3} M_{\mathbf{q}}(\mathbf{k},\mathbf{k}_1)
\chi_{\mathbf{q}}(\mathbf{k}_1)
\end{eqnarray}
where the source term,
$\chi_{\mathbf{q}}^{(0)}(\mathbf{k})$, is given by
\begin{equation}
\chi_{\mathbf{q}}^{(0)}(\mathbf{k})=
nF(|\mathbf{k}-\mathbf{q}/2|)
\frac{\tilde{v}_{\mathbf{k}+\mathbf{q}/2}(\mathbf{k}-\mathbf{q}/2,\mathbf{q})}
{m(|\mathbf{k}+\mathbf{q}/2|)}
F(|\mathbf{k}+\mathbf{q}/2|),
\end{equation}
and
\begin{eqnarray}\label{mqk}
\lefteqn{M_{\mathbf{q}}(\mathbf{k},\mathbf{k}_1) =
n\frac{F(|\mathbf{k}-\mathbf{q}/2|)}{m(|\mathbf{k}-\mathbf{q}/2|)}
\tilde{v}_{\mathbf{k}-\mathbf{q}/2}(\mathbf{k}-\mathbf{k}_1,\mathbf{k}_1-\mathbf{q}/2)}
\\ \nonumber &&  
\times
F(|\mathbf{k}-\mathbf{k}_1|)
\tilde{v}_{\mathbf{k}+\mathbf{q}/2}
(\mathbf{k}-\mathbf{k}_1,\mathbf{k}_1+\mathbf{q}/2)
\frac{F(|\mathbf{k}+\mathbf{q}/2|)}{m(|\mathbf{k}+\mathbf{q}/2|)}.
\end{eqnarray}

The linear integral operator at the right hand side of Eq. (\ref{chiqk}) 
coincides with the operator introduced by Biroli and Bouchaud \cite{BB}
and subsequently re-derived by Biroli \textit{et al.} \cite{BBMR} and 
by one of us \cite{GS4point}. This is consistent with the fact that the 
rainbow diagrams contribute to the divergence of both a three-point
susceptibility introduced by Biroli \textit{et al.} \cite{BBMR} and 
a four-point correlation function of 
Ref. \cite{GS4point}. Thus, we can conclude that re-summing rainbow diagrams
(and, more generally, re-summing two classes of diagrams showed in Fig. \ref{fig3})
amounts to adding corrections to mode-coupling theory that originate from 
critical fluctuations.

At $\mathbf{q}=0$, $M_{\textbf{0}}(\mathbf{k},\mathbf{k}_1)$
becomes the stability matrix of the mode-coupling theory \cite{Goetzebook}
(remember 
that mode-coupling expressions for $F(k)$ and $m(k)$ are used in Eq. (\ref{mqk})).
Upon approaching the mode-coupling transition its largest eigenvalue approaches 1,
\begin{equation}
\int \frac{d\mathbf{k}_1}{(2\pi)^3} M_{\mathbf{0}}(\mathbf{k},\mathbf{k}_1)
h^c(k_1)=\left(1-2g(1-\lambda)\epsilon^{1/2}\right)h^c(k),
\end{equation}
where $h^c$ is the right eigenvector of the stability matrix 
corresponding to the largest eigenvalue, 
$\epsilon$ is the fractional distance from the transition,
and $g$ and $\lambda$ are standard constants introduced in the 
mode-coupling analysis \cite{Goetzebook}.
This translates into a divergence of 
$\chi_{\mathbf{0}}$ at the transition. More generally,
close to the transition and for small $\mathbf{q}$, we have
\begin{eqnarray}\label{chiqksol}
\chi_{\mathbf{q}}(\mathbf{k}) = \epsilon^{-1/2}
\frac{n a h^c(k) S(k)}{1+\epsilon^{-1/2}\Gamma q^2}
\end{eqnarray}
where $a$ is proportional to the projection 
of the $\mathbf{q}=0$ part of the source term 
on the left eigenvector of the stability matrix  
corresponding to the largest eigenvalue \cite{Goetzebook}, $\hat{h}^c$,
\begin{eqnarray}\label{a}
a = \left(2 n g (1-\lambda)\right)^{-1}
\int_0^\infty dk \hat{h}^c(k) \chi_{\mathbf{0}}^{(0)}(k)/S(k).
\end{eqnarray}
In Eq. (\ref{chiqksol}) 
$\Gamma$ originates from the small $\mathbf{q}$ correction to the largest
eigenvalue of $M_{\mathbf{q}}(\mathbf{k},\mathbf{k}_1)$. 
$\Gamma$ can be expressed in terms of the right and left eigenvectors of the
stability matrix and equilibrium correlation functions \cite{SFH}. 

\begin{figure}
\centerline{\includegraphics[scale=.31]{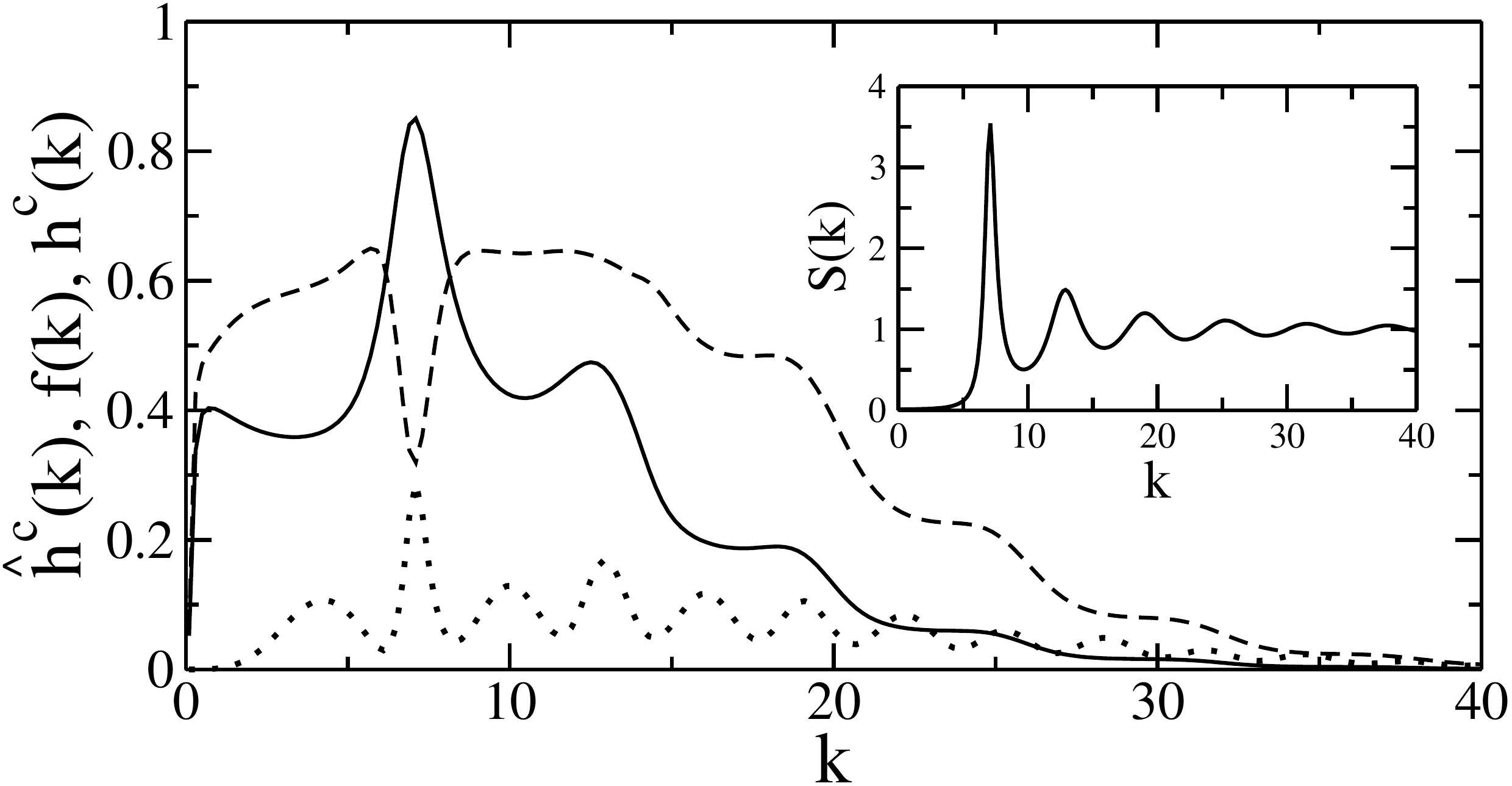}}
\caption{Wave-vector dependence of non-ergodicity parameter
$f(k)=F(k)/S(k)$ (solid line), right eigenvector of the stability matrix
$h^c(k)$ (dashed line) and left eigenvector of the stability matrix 
$\hat{h}^c(k)$ (dotted line).
Insert: static structure factor $S(k)$. All quantities calculated 
at the mode-coupling transition.}
\label{fig5}
\end{figure}

The above discussion implies that the sum of the class of diagrams showed
in the first line of Fig. \ref{fig3} reads 
\begin{eqnarray}\label{first}
\lefteqn{S(k) \int \frac{d\mathbf{k}_1 d\mathbf{q} }{(2\pi)^{6}}
\tilde{v}_{\mathbf{k}}
(\mathbf{k}_1-\mathbf{q}/2,\mathbf{k}-\mathbf{k}_1+\mathbf{q}/2)}
\\ \nonumber && \times 
\chi_{\mathbf{q}}(\mathbf{k}-\mathbf{k}_1)
\chi_{\mathbf{q}}(\mathbf{k}_1)
F(q)
\tilde{v}_{\mathbf{k}}
(\mathbf{k}_1+\mathbf{q}/2,\mathbf{k}-\mathbf{k}_1-\mathbf{q}/2).
\end{eqnarray}
Close to the transition the integral is dominated by the diverging small $\mathbf{q}$ 
contribution. In three spatial dimensions, $D=3$,
we can use asymptotic formula (\ref{chiqksol}) for all wave-vectors.  
In this way we can show that the 
singular part of (\ref{first}) is equal to $a^2 F(0) d(k)$ where $a$ is defined in
Eq. (\ref{a}) and the function $d(k)$ is given by the following formula
\begin{eqnarray}\label{firstasympt}
d(k)= \frac{n^2 S(k)}{ \epsilon^{1/4} \, 8\pi\Gamma^{3/2}}
\!\!\!\!\!\!\!\!\!\!\!\!    && \int \frac{d\mathbf{k}_1}{(2\pi)^{3}}
\tilde{v}_{\mathbf{k}}^2(\mathbf{k}_1,\mathbf{k}-\mathbf{k}_1)
\\ \nonumber && \times
S(k_1) S(|\mathbf{k}-\mathbf{k}_1|)h^c(k_1) h^c(|\mathbf{k}-\mathbf{k}_1|).
\end{eqnarray}
Explicit calculation (see Fig. \ref{fig5}) shows that $h^c(k)$ does not 
change the sign and this makes $d(k)$ a positive function. We note that
this function diverges 
as $\epsilon^{-1/4}$ upon approaching the mode-coupling transition.

The analysis of the class of diagrams showed in the second line of Fig. \ref{fig3}
follows the same line of reasoning. The final result is that close to 
the transition the singular part of the sum of the class of diagrams showed
in the second line of Fig. \ref{fig3} is equal to $b d(k)$ where the coefficient $b$
is a sum of two terms, $b=b^{(1)}+b^{(2)}$, which originate from the two parts
of the four leg-vertex. The first term reads
\begin{eqnarray}
\lefteqn{b^{(1)} = 
- \frac{1}{\left(2g(1-\lambda)\right)^2}
\int_0^\infty   \!\!\! d k_1 \frac{\hat{h}^c(k_1) F^2(k_1)}{S(k_1)}}
\\ \nonumber && \!\!\!\!\!\!\!\!\!\!\!\!  \times 
\int_0^\infty \!\!\! d k_2 \frac{\hat{h}^c(k_2) F^2(k_2)}{S(k_2)}
\int \frac{d\hat{\mathbf{k}}_1}{4\pi}
\frac{\tilde{v}_{\mathbf{k}_1+\mathbf{k}_2}^2(\mathbf{k}_1,\mathbf{k}_2)
S(|\mathbf{k}_1+\mathbf{k}_2|)}
{m(|\mathbf{k}_1+\mathbf{k}_2|)}.
\end{eqnarray}
The second term is given by a lengthier expression and it will be presented elsewhere
\cite{SFH}. 

We numerically calculated all quantities needed to evaluate the 
singular part of the total contribution, $\delta m(k) = 
\left(a^2F(0)+b\right)d(k)$.
The only 
input in this calculation is the static structure factor $S(k)$, which we calculated 
for the hard sphere interaction potential using the Percus-Yevick approximation.
We used 300 equally spaced wave-vectors with spacing 
$\delta=0.2$, between $k_0=0.1$ and $k_{\mathrm{max}}=59.9$.
For the hard sphere system the only control parameter is volume fraction 
$\varphi= n \pi \sigma^3/6$, where $\sigma$ is the hard sphere diameter. 
For our discretization the mode-coupling transition is located at 
$\varphi_c = 0.515866763$. In Fig. \ref{fig5}
we show the non-ergodicity parameter, $f(k)=F(k)/S(k)$, and 
the left and right eigenvectors, $h^c(k)$ and $\hat{h}^c(k)$, 
calculated at the mode-coupling transition. We used these functions to calculate
mode-coupling constants $g= 2.41$, $\lambda= 0.735$, and $\Gamma= 0.0708$, and 
coefficients $a^2F(0)= 1.02 \times 10^{-6}$, $b^{(1)}= - 0.1065$, and 
$b^{(2)} =  -0.07074$. We see that the contribution
of the class of diagrams showed in the second line of Fig. \ref{fig3} 
dominates and makes $\delta m(k)$ negative. 

\begin{figure}
\centerline{\includegraphics[scale=.31]{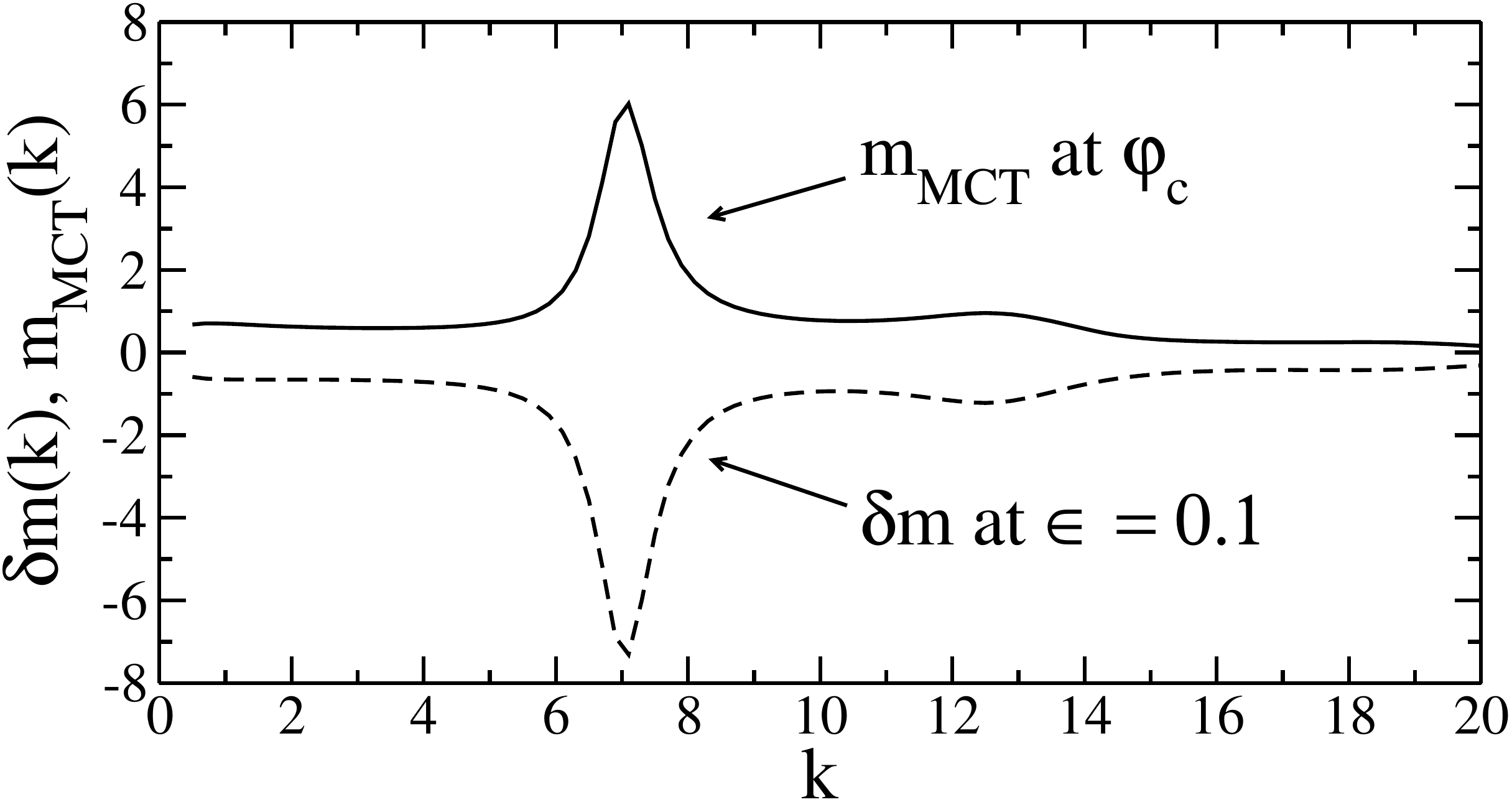}}
\caption{Mode-coupling result for the long-time
limit of the irreducible memory function, $m_{\text{MCT}}(k)$, calculated
at the transition, solid line.
Singular part of the 
contribution of two classes of diagrams showed in Fig. \ref{fig3}, 
$\delta m(k)$, calculated
at the fractional distance from the transition $\epsilon=0.1$, dashed line.}
\label{fig6}
\end{figure}

In Fig. \ref{fig6} we show the mode-coupling result for the 
long-time limit of the irreducible memory function, $m_{\text{MCT}}(k)$, 
and the singular part of the contribution from
two classes of diagrams showed in Fig. \ref{fig3}, $\delta m(k)$. 
The latter quantity is
calculated for $\epsilon = (\varphi - \varphi_c)/\varphi_c = 0.1$. We see that even at 
such a large 
$\epsilon$ 
$m_{\text{MCT}}(k)$ and $\delta m(k)$ are comparable.

On the one hand, the result showed in Fig. \ref{fig6} confirms our intuitive
expectations based on the Franz-Parisi potential picture. The sum of the contributions 
to the long-time  limit of the irreducible memory function originating from
critical fluctuations is negative and, thus, in a perturbative calculation,
the transition would be shifted towards lower temperatures and/or higher densities.
On the other hand we see that these contributions diverge at the 
transition suggesting that a perturbative expansion around mode-coupling theory 
breaks down.

Generalizing our results to higher dimensions we see that for $D>4$ the 
integral resulting from using asymptotic formula (\ref{chiqksol}) in Eq. 
(\ref{firstasympt}) is IR convergent and UV divergent. The IR contribution is
non-analytic in $\epsilon$ 
for $D<8$ and thus $D=8$ is the upper critical dimension, which agrees with 
earlier results \cite{Fetal,BBJPCM}.

\begin{figure}
\centerline{\includegraphics[scale=.135]{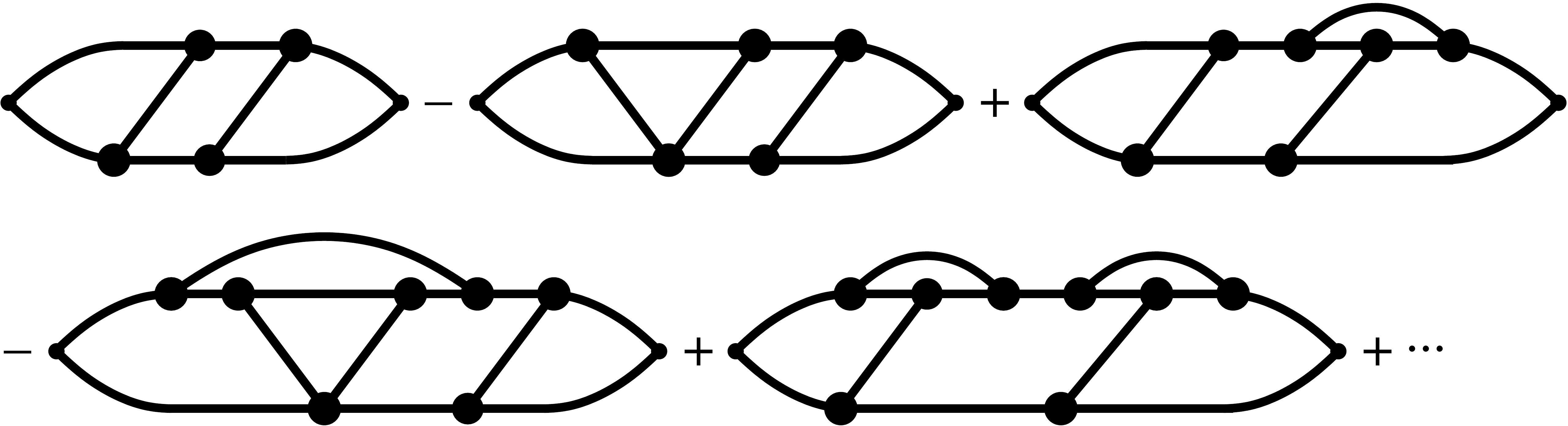}}
\caption{First few renormalized ladder 
diagrams contributing to the long-time limit of the irreducible memory function $m(k)$.
The ``rungs'' of these ladder diagrams are the same as those of diagrams showed
in Fig. \ref{fig3}. Re-summation of diagrams showed in Figs. \ref{fig3} and \ref{fig7}
results in expression (\ref{final}).}
\label{fig7}
\end{figure}
One could argue on general grounds that the full memory function $m(k)$ should be
positive, thus in addition to the negative contribution showed in Fig. \ref{fig6}
there have to be additional, positive contributions. 
In three spatial dimensions
it is possible to perform an additional re-summation of a class 
of renormalized ladder diagrams in which 
the ``rungs'' are the same as in the diagrams showed in Fig. \ref{fig3}. The 
first few ladder diagrams are showed in Fig. \ref{fig7}. One can show that 
the re-summation\footnote{Since we are considering re-summation of terms that diverge
as $\epsilon\to 0$, we perform Borel re-summation.}
of diagrams showed in  Fig. \ref{fig3} and of their ladder
counterparts showed in Fig. \ref{fig7} results in the following contribution 
to $m(k)$ 
\begin{eqnarray}\label{final}
&& \!\!\!\!\!\!\!\!\!\!\!\!
\epsilon^{-1/4} \frac{n^2 S(k) (a^2 F(0) + b)}{8\pi\Gamma^{3/2}}
\int \frac{d\mathbf{k}_1}{(2\pi)^{3}}
\tilde{v}_{\mathbf{k}}^2(\mathbf{k}_1,\mathbf{k}-\mathbf{k}_1)
\\ \nonumber && 
\times \frac{S(k_1) S(|\mathbf{k}-\mathbf{k}_1|)
h^c(k_1) h^c(|\mathbf{k}-\mathbf{k}_1|)}
{1 - 2 \epsilon^{-1/4} \frac{n(a^2 F(0) + b)}{8\pi\Gamma^{3/2}}
\frac{S(k_1) S(|\mathbf{k}-\mathbf{k}_1|)
h^c(k_1) h^c(|\mathbf{k}-\mathbf{k}_1|)}{F(k_1)F(|\mathbf{k}-\mathbf{k}_1|)}}
\end{eqnarray}
(one should remember that $(a^2 F(0) + b)<0$).
We note that as $\epsilon\to 0$ the above contribution is finite and tends to
$-m_{\text{MCT}}(k)$. Thus, it cancels the mode-coupling contribution to
$m(k)$. This fact suggests that in $D=3$ the spurious transition predicted by the
mode-coupling theory is cut off by the inclusion of critical fluctuations.

\section{Discussion} 

We showed that re-summations of two infinite classes of diagrams contributing
to the long-time limit of the irreducible memory function results in contributions that
diverge at the mode-coupling transition. The origin of the divergences are rainbow-like
diagrammatic insertions. The same insertions are responsible for
the divergence of the three-point susceptibility \cite{BBMR} and the divergent
part of a four-point correlation function \cite{GS4point}. This allows us to
associate the divergences which we identified with critical fluctuations appearing 
at the transition. Our results suggest that in three spatial dimensions
these fluctuations cut off the transition.

We emphasize that we have only considered the long-time limit 
of the irreducible memory function. In particular, we do not expect the cancellation 
that we found to also happen at finite times. We shall mention in this context a 
very recent preprint \cite{Rizzo}, which suggests, on the basis of an ingenious mapping 
of a diagrammatic expansion onto a stochastic field theory, that including all
leading order corrections cuts off the mode-coupling transition but does
not significantly change the mode-coupling predictions in the region of 
the plateaus. Most interestingly, 
according to Ref. \cite{Rizzo} the re-summation of the leading order corrections
accounts for activated dynamics.

Our fully microscopic analysis could be
compared with a recent static replica field theory investigation of critical correlations
at the dynamic transition \cite{Fetal}. 
At present, both approaches deal with different
quantities. We focused on the long-time limit of the irreducible memory function
whereas Refs. \cite{Fetal} analyzed critical correlations directly.  
Our approach could be used to investigate the three-point susceptibility of Biroli
\textit{et al.} \cite{BBMR}, which also reflects critical correlations directly. 
This is left for future research. 

Finally, from the point of view of the equation for the ergodicity breaking parameter
the long-time limit of the irreducible memory function $m(k)$ obtained from a dynamic
theory plays the same role as $nS(k)\tilde{c}(k)$ where $\tilde{c}(k)$ is the
off-diagonal direct correlation function of a static approach. Thus, our
renormalized diagrammatic expansion can be compared with a recent systematic
expansion in the non-ergodicity parameter obtained from a static replica 
approach \cite{JZ}. We note that these two
expansions are different and, thus, the relationship of dynamic and static
approaches to the glass transition appears unclear. 

\section{Acknowledgments} 

GS and EF acknowledge the support of NSF Grant CHE 1213401 and 
HH acknowledges the support of the Grant-in-Aid of MEXT (No. 25287098). 
This work was started when GS was a Visiting Professor at 
Yukawa Institute for Theoretical Physics at Kyoto University and 
was continued when he visited Department of Physics of Sapienza University
of Rome. He thanks both institutions for their hospitality.
GS acknowledges discussions with 
S. Franz, G. Parisi, T. Rizzo  and F. Zamponi.


\begin{thebibliography}{99}
\bibitem{Leutheusser} E. Leutheusser, Phys. Rev. A \textbf{29}, 2765, (1984).
\bibitem{BGS} U. Bengtzelius, W. G\"otze and A. Sj\"olander,
J. Phys. C \textbf{17}, 5915, (1984).
\bibitem{DMRT} S.P. Das, G.F. Mazenko, S. Ramaswamy and J.J. Toner,
Phys. Rev. Lett. \textbf{54}, 118 (1985).
\bibitem{Goetzebook} W. G\"otze,
\textit{Complex dynamics of glass-forming liquids: A mode-coupling theory}
(Oxford University Press, Oxford, 2008).
\bibitem{NK} M. Nauroth and W. Kob, Phys. Rev. E \textbf{55}, 657 (1997).
\bibitem{KNS} W. Kob, M. Nauroth and F. Sciortino,
J. Non-Cryst. Solids, \textbf{307}-\textbf{310}, 181 (2002).
\bibitem{GK2000} T. Gleim and W. Kob, Eur. Phys. J. B \textbf{13}, 83 (2000).
\bibitem{WPFV2010} F. Weysser, A.M. Puertas,
M. Fuchs and Th. Voigtmann, Phys. Rev. E \textbf{82}, 011504 (2010).
\bibitem{KobLH} W. Kob, in \textit{Slow Relaxations and Nonequilibrium Dynamics
in Condensed Matter}, J.-L. Barrat, M. V. Feigelman, J.
Kurchan, and J. Dalibard, eds. (Springer-Verlag, Berlin, 2003).
\bibitem{DM} S.P. Das and G.F. Mazenko, Phys. Rev. A \textbf{34}, 2265 (1986).
\bibitem{GScutoff} W. G\"otze and L. Sj\"ogren, Z. Phys. B \textbf{65}, 415 (1987).
\bibitem{CR} M.E. Cates and S. Ramaswamy, Phys. Rev. Lett. \textbf{96}, 135701 (2006).
\bibitem{SE} G. Szamel and E. Flenner, Europhys. Lett. \textbf{67}, 779 (2004).
\bibitem{FP} S. Franz and G. Parisi, J. Phys. I France \textbf{5}, 
1401 (1995); Phys. Rev. Lett. \textbf{79}, 2486 (1997).
\bibitem{CFP}
M. Cardenas, S. Franz and G. Parisi, J. Chem. Phys. \textbf{110}, 1726 (1999).
\bibitem{Franz} S. Franz, Europhys. Lett. \textbf{73} 492 (2006).
\bibitem{FM} S. Franz and A. Montanari, J. Phys. A: Math. Theor. 40 F251 (2007).
\bibitem{Ken} K.S. Schweizer and E.J. Saltzman, J. Chem. Phys. \textbf{119}, 1181 (2003);
K.S. Schweizer, J. Chem. Phys. \textbf{123}, 244501 (2005).
\bibitem{Fetal} S. Franz, H. Jacquin, G. Parisi, P. Urbani and F. Zamponi, 
Proc. Natl. Acad. Sci. U.S.A. \textbf{109}, 18725 (2012); \textit{id.}, 
J. Chem. Phys. \textbf{138}, 12A540 (2013).
\bibitem{gMCT} G. Szamel, Phys. Rev. Lett. \textbf{90}, 228301 (2003).
\bibitem{WC} J. Wu and J. Cao, Phys. Rev. Lett. \textbf{95}, 078301 (2005).
\bibitem{Mayer} P. Mayer, K. Miyazaki and D.R. Reichman, 
Phys. Rev. Lett. \textbf{97}, 095702 (2006).
\bibitem{ABL} A. Andreanov, G. Biroli and A. Lefevre, J. Stat. Mech.: Theory Exp. 
(2006) P07008.
\bibitem{KK} B. Kim and K. Kawasaki, J. Phys. A \textbf{40}, F33 2007 ; J. Stat.
Mech.: Theory Exp. (2008) P02004.
\bibitem{NH} T.H. Nishino and H. Hayakawa, Phys. Rev. E \textbf{78}, 061502 (2008).
\bibitem{JW} H. Jacquin and F. van Wijland, Phys. Rev. Lett. \textbf{106}, 210602 (2011).
\bibitem{GSdiagram} G. Szamel, J. Chem. Phys. \textbf{127}, 084515 (2007).
\bibitem{CHess} B. Cichocki and W. Hess, Physica A \textbf{141}, 475 (1987).
\bibitem{GS4point} G. Szamel, Phys. Rev. Lett. \textbf{101}, 205701 (2008).
\bibitem{GSPTEP} G. Szamel, Prog. Theor. Exp. Phys. 012J01 (2013).
\bibitem{SFH} G. Szamel, E. Flenner and H. Hayakawa, in preparation.
\bibitem{SL} G. Szamel and H. L\"{o}wen, Phys. Rev. A \textbf{44}, 8215 (1991).
\bibitem{BB} G. Biroli and J.-P. Bouchaud, Europhys. Lett. \textbf{67}, 21 (2004).
\bibitem{BBMR} G. Biroli, J.-P. Bouchaud, K. Miyazaki, and D. R. Reichman,
Phys. Rev. Lett. \textbf{97}, 195701 (2006).
\bibitem{BBJPCM} G. Biroli and J.-P. Bouchaud, J. Phys.: Condens. Matter \textbf{19},
205101 (2007).
\bibitem{Rizzo} T. Rizzo, arXiv:1307.4303.
\bibitem{JZ} H. Jacquin and F. Zamponi, J. Chem. Phys. \textbf{138}, 12A542 (2013).


\end{thebibliography}
\end{document}